\documentclass[aps,prb,twocolumn]{revtex4}
\usepackage{epsf,graphicx}
\usepackage{amssymb}
\usepackage{amsmath}
\usepackage{eepic}

\newlength{\piclen}
\begin{document}

\title{Origin of large thermopower in LiRh$_2$O$_4$}

\author{R.\ Arita$^{1,*}$, K.\ Kuroki$^2$, 
K.\ Held$^3$, A.\ V.\ Lukoyanov$^4$, S.\ Skornyakov$^5$,
V.\ I.\ Anisimov$^5$}
\affiliation{$^1$RIKEN (The Institute of Physical and Chemical Research),
Wako, Saitama 351-0198, Japan\\
$^2$ University of Electro-Communications 1-5-1 Chofugaoka, Chofu-shi
Tokyo 182-8585, Japan\\
$^3$ Institute for Solid State Physics, Vienna University of Technology,
1040 Vienna, Austria \\
$^4$ Ural State Technical University-UPI, 620002 Yekaterinburg, Russia\\
$^5$ Institute of Metal Physics, Russian Academy of Science-Ural division,
620219 Yekaterinburg, Russia
}

\date{\today}

\begin{abstract}
Motivated by the newly synthesized mixed-valent
spinel LiRh$_2$O$_4$ for which a large thermopower
is  observed in the  metallic  cubic phase above 230K
[Okamoto {\it et al.}, Phys. Rev. Lett. {\bf 101}, 086404(2008)], 
we calculate
the Seebeck coefficient by the combination of local density 
approximation and dynamical mean field theory (LDA+DMFT).
The experimental values are well reproduced not
only by LDA+DMFT but also by the less involved
Boltzmann equation approach. A careful analysis of
the latter shows unexpectedly  that the origin of the large thermopower 
shares a common root with a very different oxide: Na$_x$CoO$_2$. We also
discuss how it is possible to further increase the power factor of 
LiRh$_2$O$_4$ through doping, which makes
the material even more promoising for technological applications.

\end{abstract}

\pacs{71.15.Jf, 71.15.-m}

\maketitle

\section{Introduction}
Designing and searching for good thermoelectric 
materials have a long history of extensive studies
due to the scientific interest and potential
technological importance, particularly 
for generating electrical power from heat 
(gradients) and for cooling through the  Peltier effect \cite{Mahan}.
Hitherto, the main target materials have been various
insulators or semiconductors such as Bi$_2$Te$_3$ \cite{Mahan} and 
FeSb$_2$\cite{FeSb2}, since it was believed
that huge thermopowers cannot be expected for metals. 
However, recently, novel metallic systems
with large thermopower have been discovered and 
attracted much attention. Generally, materials
with  strong electronic correlations are promising \cite{Paschen};
and a  famous example is Na$_x$CoO$_2$,
for which a metallic resistivity as low as $\rho$=0.2 m$\Omega$cm
and a  thermopower as large as
$S$=100 $\mu$V/K are observed
simultaneously at 300K \cite{Terasaki}. 
The coexistence of low resistivity and 
large thermopower results in a large power factor 
($S^2/\rho$), which is especially important for device applications.

Most recently, Okamoto {\it et al.} 
synthesized a new mixed-valent spinel oxide,
LiRh$_2$O$_4$\cite{Okamoto}. 
This novel oxide shows two structural phase 
transitions, i.e., the cubic-to-tetragonal transition 
at 230K and the tetragonal-to-orthorhombic transition at
170K.  Particularly interesting is  however the high temperature
cubic phase: Despite the metalicity
which is reflected in a  small resistivity 
and  the existence of a Fermi-edge,
the thermopower is as large as $80 \mu$V/K at 800K, 
which is exceptional for metallic systems.

On the theoretical side, a variety of
studies have been performed to understand the mechanism
of large thermopowers in  metallic systems. 
Among others, Koshibae {\it et al.} derived an expression 
for the Seebeck coefficient of strongly correlated
systems in the high-temperature limit \cite{Koshibae}. 
Considering the orbital and spin degrees 
of freedom of localized electrons, they estimated the thermoelectric 
power of Na$_x$CoO$_2$ to be 150 $\mu$V/K. 

However, when the temperature ($T$) is much lower than the
energy scale of the band width ($\sim 2$ eV),
it is expected that the band 
dispersion of the system also plays a crucial role
as has been suggested from first principle (bandstructure)
 studies \cite{Singh,Wilson}.
Indeed, recently, two of the present authors
proposed that the peculiar shape of the valence band 
(the so-called $a_{1g}$ band) is important to realize a
large thermopower and high conductivity in Na$_x$CoO$_2$\cite{Kuroki}.
The different theoretical proposals led to a heated discussion
\cite{hotdebate} and also to a proposal to discriminate between
them through the respective temperature dependence \cite{Ishida}. 

The motivation of the present study is 
to clarify the origin of the large thermopower in LiRh$_2$O$_4$.
For this purpose, we first perform a LDA+DMFT \cite{LDA+DMFT}
calculation (the combination of the local density 
approximation and the dynamical mean field theory\cite{DMFT}), 
employing the Kubo formula for the Seebeck 
coefficient \cite{Oudovenko}. This {\em ab initio} approach is
going way beyond  Ref.\ \onlinecite{Kuroki} where several phenomenological 
parameters had to be introduced.
Second, we study whether the Boltzmann equation approach with
the  LDA band dispersion as an input works well for this system. We will show
that this approach gives results quantitatively similar to those of LDA+DMFT.
Even though LiRh$_2$O$_4$ is a material very different from Na$_x$CoO$_2$,
having among others a much more complicated bandstructure,
our analysis nontheless  reveals that the origin of the large 
thermopower is similar:
the ``pudding mold'' shape of the bands crossing the Fermi energy.
This outcome was  not prejudiced in our investigation
and is quite surprising.
We also discuss how electron doping could further increase the 
power factor of LiRh$_2$O$_4$.

\section{Method}
As a first step we do a LDA calculation for LiRh$_2$O$_4$,
using the linearized muffin tin orbital (LMTO) basis set \cite{LMTO},
employing the experimental  lattice constant $a=8.46$
and  so-called $x$ parameter $x=0.261$ (which indicates the position 
of the oxygen sites). 

\begin{figure}[tb]
\begin{center}
\includegraphics[width=8.5cm]{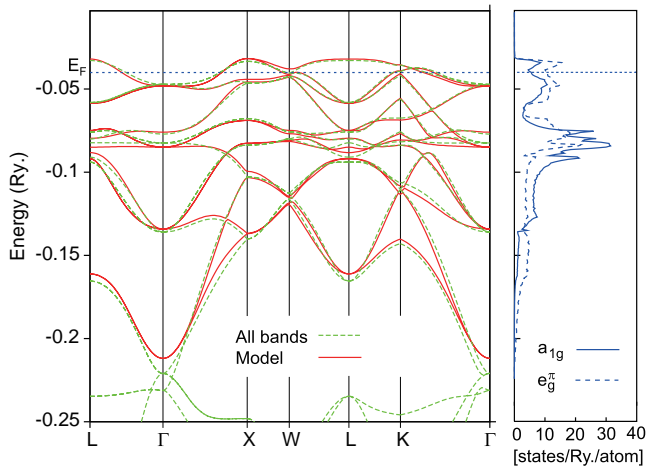}
\end{center}
\caption{Left panel:
(Color online) Band dispersion of the effective 3-orbital Hamiltonian
(solid line) and total LMTO band structure (dashed line) of LiRh$_2$O$_4$.
Right panel: partial $a_{1g}$ and $e_g^\pi$ density of states for the model.
}
\label{Fig:LDA}
\end{figure}

From the LMTO bandstructure, we construct an effective Hamiltonian 
($\equiv H^{\rm LDA}_{\alpha\beta})$
by the projection onto Wannier functions \cite{Projection}.
Since the unit cell of LiRh$_2$O$_4$ contains 
four Rh atoms and each Rh atom has three $t_{2g}$ orbitals
the size of the effective Hamiltonian is 12$\times$12. 
A comparison of the band dispersion of this effective Hamiltonian 
with the total LDA band structure is shown in Fig.\ \ref{Fig:LDA}.
In contrast to the case of
Na$_x$CoO$_2$ \cite{Kuroki}, not only the $a_{1g}$-orbital
but also the $e_{g}^{\pi}$-orbitals have a  substantial density
of states (see the right panel of Fig.\ref{Fig:LDA}) 
at the Fermi level ($E_F$). 
Hence, we cannot extract a 
simpler effective Hamiltonian from
the 12 $\times$ 12 Hamiltonian and need to
keep all $t_{2g}$ orbitals in the following calculation.

Next, we supplement the 3-orbital Hamiltonian
by local intra- ($U$) and inter-orbital ($U'$)
Coulomb repulsions as well as by Hund's exchange ($J$; of Ising type),
and solve it by DMFT\cite{DMFT}, 
using the quantum Monte Carlo (QMC) method. 
To get high-quality QMC data, we take $\sim 3.0\times 10^7$ 
sweeps  in the calculation.
%since simulating the SU(2) symmetric Hund coupling is difficult
%in QMC. 
%The application of new, more
%sophisticated algorithms to this end, such as \cite{Sakai},
%remains an important challenge for the future.

In the framework of DMFT, the Kubo formula for the 
Seebeck coefficient is \cite{Oudovenko},
\begin{equation}
S=\frac{k_B}{e}\frac{A_1}{A_0},
\label{seebeckeq1}
\end{equation}
where $k_B$ and $e$ are the Boltzmann constant and unit charge, respectively,
and
\begin{eqnarray}
A_n&=&2\pi\hbar \int^{\infty}_{-\infty}d\omega
\phi^{xx}(\omega)f(\omega)f(-\omega)(\beta \omega)^n, \\
\phi^{xx}(\omega)&=&\frac{1}{V}\sum_{\bf k} {\rm Tr} \left(
v^x({\bf k})\rho({\bf k},\omega)
v^x({\bf k})\rho({\bf k},\omega)
\right).
\label{seebeckeq2}
\end{eqnarray}
Here, $\rho({\bf k},\omega)$ is the spectral function, i.e.,
the imaginary part of the Green function $G({\bf k},\omega)$;
$v_{\alpha\beta}({\bf k})\equiv
\langle k\beta|(1/m)\nabla_x|k\alpha\rangle$
is the  group velocity, $f(\omega)$
the Fermi-Dirac distribution function, and
$V$ the volume of the unit cell.

As is carefully discussed in Ref.\ \onlinecite{Paul},
when the tight-binding basis is well localized in
the real space, we can use the so-called Peirls approximation,
$v_{\alpha\beta}({\bf k})=
\nabla_{\bf k} H^{\rm LDA}_{\alpha\beta}({\bf k})$\cite{Oudovenko}. 
In this method, since we have an analytical expression of 
$H^{\rm LDA}_{\alpha\beta}({\bf k})$, the mesh for
the momentum sum in Eq.(\ref{seebeckeq2}) can be
arbitrarily dense. In most cases we took a
 40$\times$40$\times$40 mesh, but 
in some cases also a 80$\times$80$\times$80 mesh for
checking convergence.

Usually,
 $G({\bf k},\omega)$ is calculated in DMFT(QMC)
from the self energy $\Sigma(\omega)$ which is
obtained as a root from the local Green function
 $G^{\rm imp}(\omega)$, obtained in turn 
from the QMC data by
 the maximum entropy method (see, e.g., 
Refs.\ \onlinecite{Pchelkina,Nekrasov}).
However, this standard approach does not work  well
for the calculation of the  Seebeck 
coefficient because of the following:
Since  $\phi^{xx}(\omega)$ only
 contributes to  $A_m$ for $|\omega|\leq k_B T$,
we need  $\Sigma(\omega)$ for small $|\omega|$. 
For such frequencies  $\Sigma(\omega)$ is quite small
(smaller than 0.1eV for $|\omega|\leq k_B T$, see below).
As is pointed out in Ref.\onlinecite{Oudovenko},
this smallness makes it difficult to calculate 
 $\Sigma(\omega)$ reliably especially by a
probability-based algorithms such as the maximum entropy method.

Hence, in the present study, we calculate $\Sigma(\omega)$ directly
from $\Sigma(i\omega)$, using both the Pade approximation
and a polynomial fit. For the former, we apply the algorithm
proposed in Ref.\ \onlinecite{Pade} to the data with $i\omega\in [0,45i]$eV.
For the latter, we fit $\Sigma(i\omega)$ for 
$i\omega\in [0,4i]$eV to $\sum_{n=0}^5 c_n (i\omega)^n$ by a
standard least-squares fit. 
Since only the behavior at small $|\omega|$ is relevant for 
the Seebeck coefficient, we can expect
the polynomial fit to give reasonable results. 
The Pade approximation might become 
problematic if   poles are present in the vicinity of 
the real-$\omega$ axis. However,
as we will see below, the resulting $\Sigma(\omega)$ for
the present case does not show any anomalous behavior for 
small $|\omega|$, which implies that the Pade approximation
is not  problematic.

\begin{figure}[tb]
\begin{center}
\includegraphics[width=8.5cm]{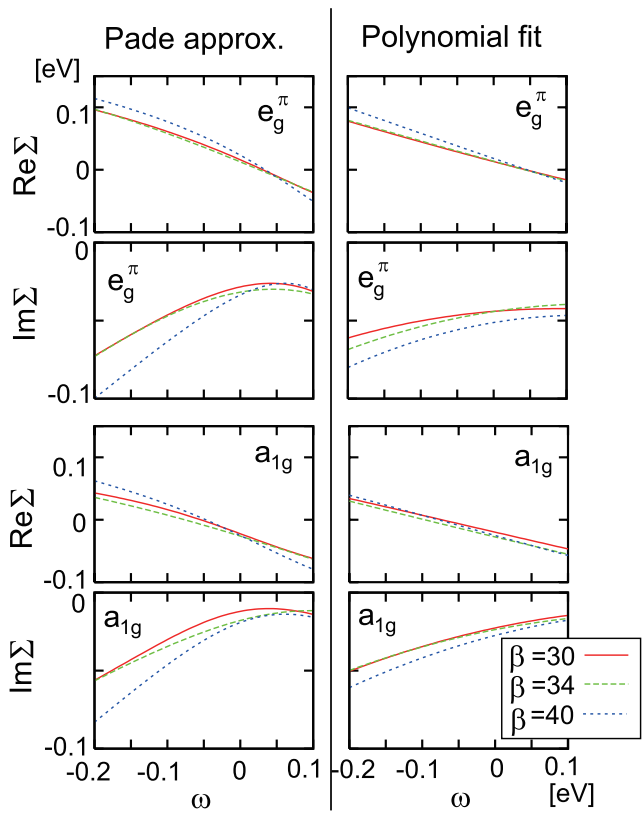}
\end{center}
\caption{(Color online) LDA+DMFT(QMC)
self energy calculated by the Pade approximation (left)
and a polynomial fit (right).
}
\label{Fig:Sigma}
\end{figure}

In Fig. \ref{Fig:Sigma},
we plot $\Sigma(\omega)$ for $(U, U',J)=(3.1, 1.7, 0.7)$eV  
which was estimated in Ref.\  \onlinecite{Pchelkina} and
$\beta=1/k_BT=$30, 34, 40 eV$^{-1}$. 
For  $T\sim 300$K, the main contribution stems from 
$\omega\in [-0.03,0.03]$eV. For these energies
 Pade approximation and
 polynomial fit  give similar results.
Even though the agreement is not perfect,
differences are small, i.e., of $O(0.01)$ eV.
Thus we employ  $\Sigma(\omega)$ of both, Pade approximation
and polynomial fitting, in the following  LDA+DMFT calculation
of the Seebeck coefficient. The difference gives us
an estimate for the accuracy of our calculation.

Besides the LDA+DMFT study, we also performed  calculations based
on the Boltzmann equation.
The Seebeck coefficient can be estimated by calculating
\begin{eqnarray}
S&=&\frac{1}{eT}\frac{K_1}{K_0}, \\
K_n&=&\sum_{{\bf k},\alpha} \tau u_\alpha({\bf k})u_\alpha({\bf k})
\left(-
\frac{\partial f(\epsilon)}
{\partial \epsilon}
\right)_{\epsilon=\epsilon({\bf k})_\alpha}
\epsilon({\bf k})_\alpha^n.
\end{eqnarray}
Here, $\tau$ is the relaxation time which we assume to be
independent of ${\bf k}$;
$\epsilon_{\alpha}({\bf k})$ are the eigenvalues of 
$H^{\rm LDA}_{\alpha,\beta}({\bf k})$; and 
$u_{\alpha}({\bf k})$  the diagonal elements of 
$\tilde{U}^{\dagger}v_{\alpha,\beta}({\bf k})\tilde{U}$, 
where $\tilde{U}$ is the unitary transformation which 
diagonalizes $H^{\rm LDA}_{\alpha\beta}({\bf k})$.

Note that $K_n$ can be roughly estimated as
\begin{eqnarray}
K_0\sim{\tilde \sum}(u_A^2+u_B^2), \\
K_1\sim(k_BT) {\tilde \sum}(u_B^2-u_A^2),
\label{seebeckeq3}
\end{eqnarray}
apart from a constant factor\cite{Kuroki}. 
Here, ${\tilde \sum}$ is a summation over the states 
in the range of $|\epsilon(k)|<O(k_BT)$,
and $u_A$ and $u_B$ are typical velocities for the states
below and above the Fermi level, respectively.

\section{Results}
\begin{figure}[t]
\begin{center}
\includegraphics[width=8.0cm]{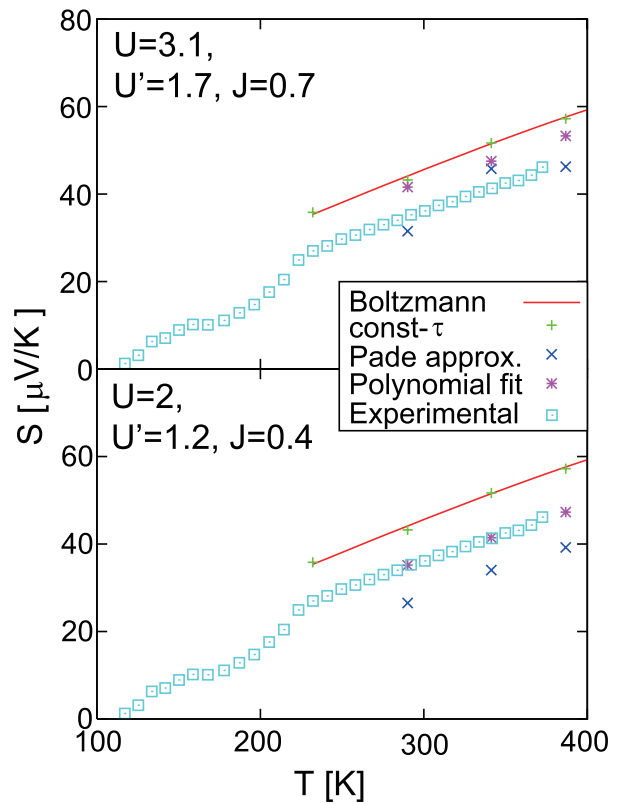}
\end{center}
\caption{(Color online)
Thermopower calculated by 
the Boltzmann equation approach and the constant-$\tau$ method 
as well as by LDA+DMFT, using both
the Pade approximation and a polynomial fit for the self energy.
}
\label{Fig:S}
\end{figure}

In Fig.\ \ref{Fig:S}, we show the resulting Seebeck coefficient
calculated by the LDA+DMFT method and the Boltzmann equation
approach. We also plot the result of the constant-$\tau$ 
approximation for the Kubo formula, i.e., 
we assume $\Sigma(\omega)=-1.0^{-3}i$ for
Eqs.\ (\ref{seebeckeq1})-(\ref{seebeckeq2}).

From Fig.\ \ref{Fig:S}, we  see that 
(1) the Boltzmann equation and the constant-$\tau$ approximation
for the Kubo formula give almost the same result;
(2) the constant-$\tau$ approximation gives a larger thermopower
than LDA+DMFT;
(3) this smaller LDA+DMFT thermopower is closer to
experiment  \cite{Okamoto}, already  
for ($U, U', J$)=(3.1,1.7,0.7)eV \cite{Pchelkina}  but even more
so for  somewhat smaller values of the Coulomb interaction.

Point (1) demonstrates that the calculation via 
Eq. (\ref{seebeckeq1}) is working well, 
if $\Sigma(\omega)$ is correct. 
Point (2)  can be understood from the behavior of
$-{\rm Im}\Sigma(\omega)$:
Fig.\ \ref{Fig:Sigma}, shows
that $-{\rm Im}\Sigma(\omega)$ calculated
by the LDA+DMFT method
is large for negative $\omega$ but small
for positive $\omega$ (independently of
Pade approximation and polynomial fit).
This means that, in contrast to the constant-$\tau$ approximation,  
the actual life time of quasi-holes
is longer than that for quasi-particles.
Therefore, the contribution of the quasi-holes (particles) to
$\phi^{xx}$ in Eq.(\ref{seebeckeq1}) becomes
larger (smaller) in the LDA+DMFT calculation, 
and consequently the first moment, $A_1$, becomes smaller.
Here, it should be noted that the constant-$\tau$ approximation
does not correspond to the limit of $U=U'=J=0$, since
this asymmetry of life time exists even in the weak coupling limit.
This is the reason why the results of LDA+DMFT move away from those 
of the constant-$\tau$ approximation as $U$, $U'$ and $J$ are decreased.

As for point (3), we would like to note
that the correlations renormalize
 the bandwidth. This renormalization is calculated
microscopically here whereas it has
been   adjusted  to the  angle-resolved
photoemission spectrum in Ref.\ \onlinecite{Kuroki}.

\section{Discussion}
While there are some differences between 
 Boltzmann equation approach and LDA+DMFT, the
results are  still very similar, even {\it quantitatively}. Hence,
we may expect that, in the present case, the Boltzmann equation can be used as
a convenient tool to analyze the mechanism of 
the large thermopower, or even to design more efficient
thermoelectric materials.

\begin{figure}[tb]
\begin{center}
\includegraphics[width=8.5cm]{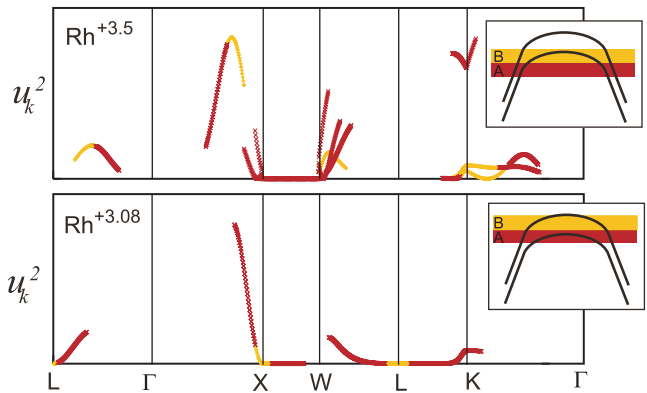}
\end{center}
\caption{(Color online)
Group velocity squared ($u_k^2$) along different 
directions of the fisrt Brillouin zone for Rh$^{+3.5}$ (LiRh$_2$O$_4$; upper panel) and Rh$^{+3.08}$ (electron-doped LiRh$_2$O$_4$; lower panel). 
$k$ point above the Fermi energy $E_F$
are shown in yellow, those below $E_F$ in red.
}
\label{Fig:vel2all}
\end{figure}

Let us first examine whether the mechanism proposed for 
Na$_x$CoO$_2$ in Ref. \onlinecite{Kuroki} can work also in LiRh$_2$O$_4$.
If the valence band has a peculiar shape of dispersion which is
dispersive below the $E_F$ but 
somewhat flat above (the so-called ``pudding-mold'' type),
$K_1$ in Eq.\ (\ref{seebeckeq3}) becomes large, since 
the group velocity above $E_F$ ($u_A^2$) is much larger 
than the one below  $E_F$ ($u_B^2$) in this case.
This is the basic idea of  Ref. \onlinecite{Kuroki} how to realize 
a large thermopower and a
low resistivity at the same time\cite{comm2}. 
In the top panel of Fig. \ref{Fig:vel2all}, we plot 
the group velocity squared for LiRh$_2$O$_4$ 
within the energy window of $|\varepsilon-E_F|<3k_BT$ at $T\simeq 300$K. 
We see that $u_A^2$ is indeed larger than $u_B^2$, confirming
this mechanism.
We note here that although 
the Rh valence is +3.5 in LiRh$_2$O$_4$, the degeneracy of 
$d_{xy}$, $d_{yz}$ and $d_{zx}$ orbitals in the cubic phase 
makes the number of holes {\it per band} small, 
resulting in a situation similar to Na$_x$CoO$_2$  
with the Co valence smaller than +3.5. 
This view is consistent with the experimental 
fact that the thermopower is suppressed in the tetragonal 
phase below 230K, where the degeneracy is lifted.\cite{Okamoto}

However, we  also see that the (squared) group velocity
above $E_F$ is still  large
for some $k$-points. In fact, for LiRh$_2$O$_4$,
there are two pudding mold bands. For the Rh valence of +3.5,
$E_F$ lies near the bending point of
one of the pudding mold bands, but also cuts through the dispersive
portion of the other (see the upper inset of Fig. \ref{Fig:vel2all}).
The former enhances the thermopower,
while the latter suppresses it. This might be the reason why
LiRh$_2$O$_4$ is not such a good  thermoelectric material 
as Na$_x$CoO$_2$.

\begin{figure}[tb]
\begin{center}
\vspace{.2cm}
\includegraphics[width=8.5cm]{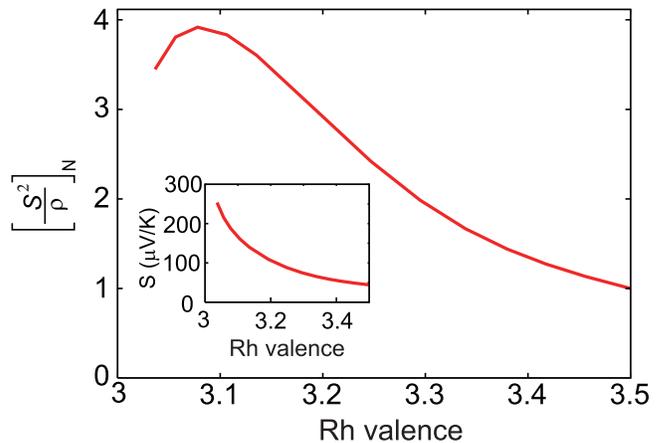}
\end{center}
\caption{(Color online)
Power factor (normalized by its value at Rh valence=+3.5 )
and thermopower (inset) as a function
of the valence of Rh, calculated by the Boltzmann equation. 
}
\label{Fig:val-seeb-pwf0026}
\end{figure}

To enhance the thermopower in LiRh$_2$O$_4$, we suggest
the following possibility:
If we  dope electrons to this system,
one of the pudding mold bands will be brought completely 
below $E_F$  (see the lower inset of Fig. \ref{Fig:vel2all}). 
The second  panel of Fig. \ref{Fig:vel2all}
shows the group velocity squared for such
a doped system with Rh valence +3.08.
In this case, the group velocity squared above $E_F$ 
is small for the entire Brillouin zone.

To confirm this idea, we calculate the thermopower and 
the power factor (normalized by its value at Rh valence=+3.5)
for various Rh valences by means of
the Boltzmann equation approach \cite{comment}, see
Fig. \ref{Fig:val-seeb-pwf0026}.
The results indicate a maximal power factor ($=S^2/\rho\propto K_1^2/K_0^3$) 
at a valency of  +3.08, where it is almost
four times larger than for LiRh$_2$O$_4$.
While the orbital degeneracy of $d_{xy}$, $d_{yz}$
and $d_{zx}$ already plays a crucial role to make $E_F$ be higher
than those of single-orbital systems\cite{Okamoto},
realizing this situation experimentally
is an interesting challenge which seems to be feasible.

%{\em In conclusion,}
%we studied the large thermopower observed in the
%high temperature cubic phase of LiRh$_2$O$_4$, using  
%the LDA+DMFT method and the Boltzmann equation approach.
%Our results suggest that the origin of the
%large thermopower is the peculiar ``pudding-mold'' shape of the valence band,
%and we predict an even higher power factor
%for the electron-doped system.

\section{Acknowledgment}
We would like to thank H. Takagi and Y. Okamoto 
for fruitful discussions and providing the experimental data
shown in Fig.\ref{Fig:S}.
Numerical calculations were performed at
the facilities of the Supercomputer center,
ISSP, University of Tokyo.
This work was supported by Grants-in-Aid
for Scientific Research (MEXT Japan) grant
19019012,19014022, 19051016
and Russian Foundation for Basic Research (RFBR) grant 07-02-00041.


\begin{references}
\bibitem{comment} Note that this
calculation neglects the  energy  and filling dependence  
of $\tau$, which should be present and affect $\rho$ and the power factor. 
Nevertheless, since the effect of the frequency dependence of $\tau$ 
is not so serious in the calculation of the Seebeck 
coefficient (Fig.\ \ref{Fig:S}),
we believe that the present estimation of the power factor 
at least gives the correct tendency of the filling dependence, 
especially if the Rh valence does not change drastically.
\end{references}
\end{document}